\documentclass[12pt]{book}

\usepackage[dvips]{graphicx,color}
\usepackage{makeidx,tsukuba}

\makeauthorindex
\makeindex

\begin{document}
\BookTitle{\itshape The 28th International Cosmic Ray Conference}
\CopyRight{\copyright 2003 by Universal Academy Press, Inc.}
%\tableofcontents
\pagenumbering{arabic}
\chapter{The AMS-02 RICH Imager Prototype \\ 
In-Beam Tests with 20 GeV/$c$ per Nucleon Ions}
\author{ M. Bu\'enerd,$^2$ on behalf of the AMS-RICH collaboration: \\
P. Aguayo,$^4$ M. Aguilar Benitez,$^4$ L. Arruda,$^3$ F. Barao,$^3$ A. Barrau,$^2$ 
B. Baret,$^2$ E. Belmont,$^6$ J. Berdugo,$^4$ G. Boudoul,$^2$ J. Borges,$^3$ 
D. Casadei,$^1$ J. Casaus,$^4$ C. Delgado,$^4$ C. Diaz,$^4$
L. Derome,$^2$ L. Eraud,$^2$ L. Gallin-Martel,$^2$ F. Giovacchini,$^1$ 
P. Goncalves,$^3$ E. Lanciotti,$^4$ G. Laurenti,$^1$ A. Malinine,$^5$ C. Mana,$^4$ 
J. Marin,$^4$ G. Martinez,$^4$ A. Menchaca-Rocha,$^6$ C. Palomares,$^4$  
M. Pimenta,$^3$ K. Protasov,$^2$ E. Sanchez,$^4$ E-S. Seo,$^5$
I. Sevilla,$^4$ A. Torrento,$^4$ M. Vargas-Trevino$^2$ \\
{\it (1) University of Bologna and INFN, Via Irnerio 46, I-40126 Bologna, Italy \\
(2) LPSC, Avenue des Martyrs 53, F-38026 Grenoble-cedex, France \\
(3) LIP, Avenida Elias Garcia 14-1, P - 1000 Lisboa, Portugal \\
(4) CIEMAT, Avenida Complutense 22, E-28040, Madrid, Spain \\
(5) U. Maryland, College Park MD 20742, USA \\
(6) Instituto de Fisica, UNAM, AP 20-364, Mexico DF, Mexico} \\}%% end of author
\section*{Abstract} 
{\small 
A prototype of the AMS Cherenkov imager (RICH) has been tested at CERN by means of 
a low intensity 20~GeV/$c$ per nucleon ion beam obtained by fragmentation of a primary beam 
of Pb ions. Data have been collected with a single beam setting, over the range of 
nuclear charges 2~$<Z<\;\sim$45 in various beam conditions and using different radiators.
The charge $Z$ and velocity $\beta$ resolution have been measured.}
\section{Introduction}
The AMS spectrometer will be installed on the International Space Station in the year 2005, for
a several years campaign of measurements during which a research program of fundamental interest 
will be covered: Search for antimatter of primordial origin and for dark matter, including
also the study of the Cosmic Ray flux (CR) and gamma ray astronomy. 

The Cherenkov imager of the experiment should make possible a thorough study of the nuclear 
Cosmic Ray flux by performing: a) Isotope identification (ID) for the light elements over a 
range of mass ($A$) and momentum of about ($A<\;\sim$15--20, 
1~GeV/$c$~$<\frac{P}{A}<\;\sim$12~GeV/$c$), 
and b) Charge measurements of the particles with charge resolution of the order of one unit around 
Fe ($Z<\,\sim$26), the momentum range extending up to the upper limit of the spectrometer 
capability, in the TeV/$c$ per nucleon range, $\frac{P}{A}<\;\sim$1~TeV/$c$, for the latter.

The technical requirements constraining the counter design (dimensions, weight, power consumption, 
long term reliability of photodetectors) have lead to the design of a proximity focusing type of 
imager, equipped with multianode photomultipliers for photon detection. The radiators were chosen in 
accordance with the physics requirements and ID capabilities of the counter: Sodium fluoride (NaF) 
and silica aerogel, to cover the low (1~GeV/$c$~$<\frac{P}{A}<\;\sim$5~GeV/$c$) and high 
(4~GeV/$c$~$<\frac{P}{A}<\;\sim$12~GeV/$c$) parts of the momentum range, respectively. 
%[*6SIMU].
%
\section{Counter architecture}
\begin{figure}[t]
  \begin{center}
    \includegraphics[width=5cm]{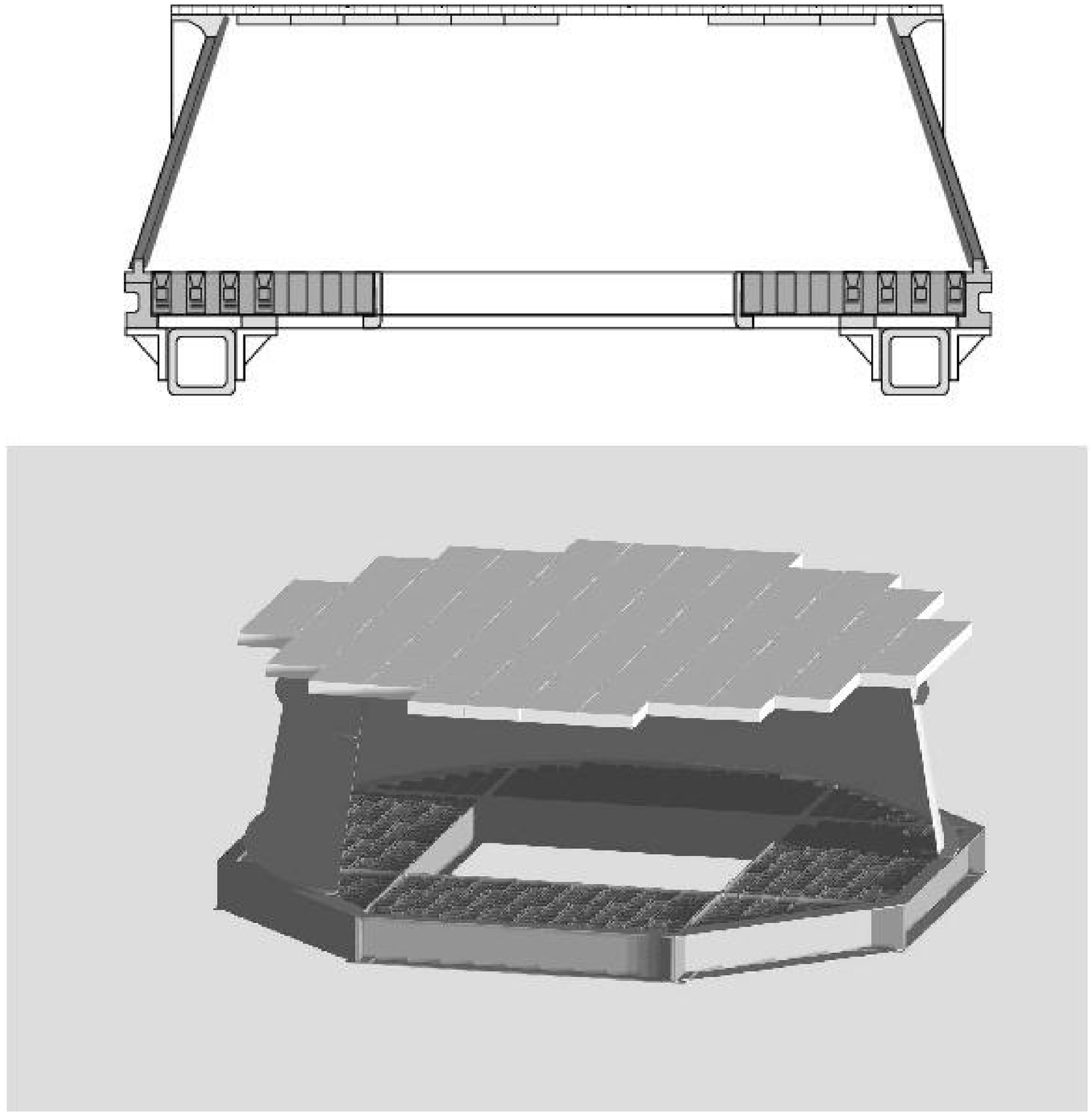}
    \includegraphics[height=5cm]{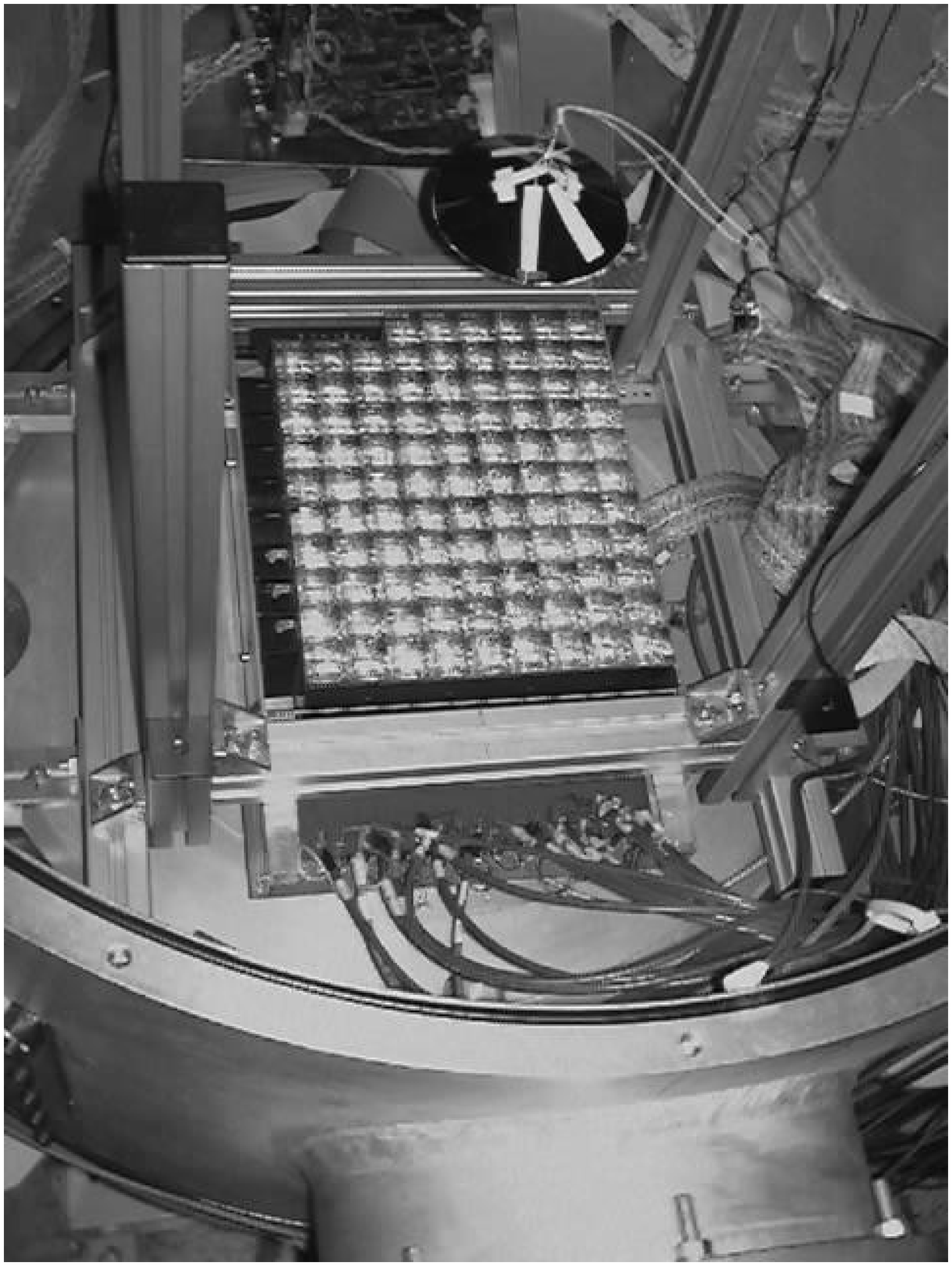}
    \includegraphics[width=5cm]{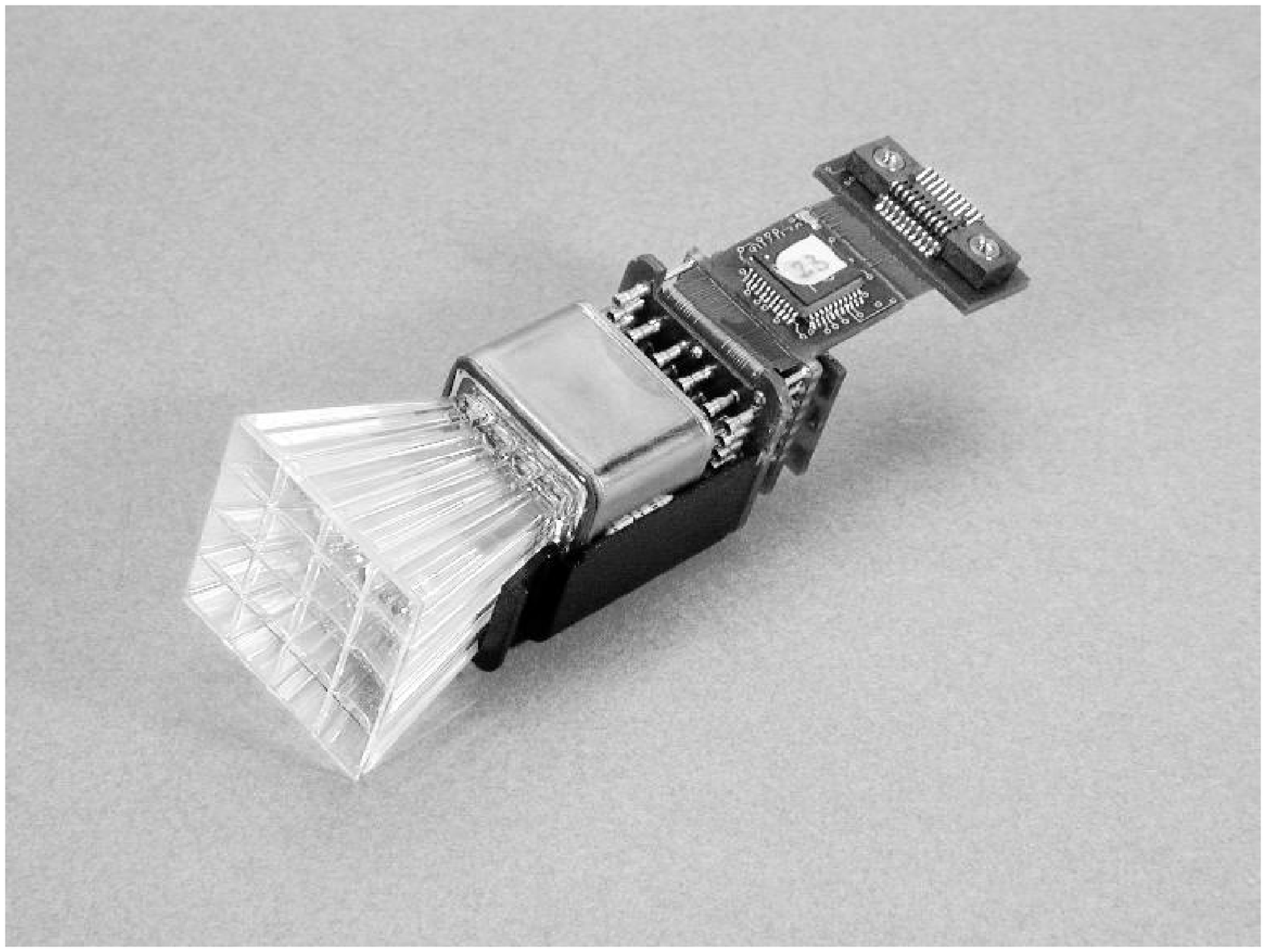}
  \end{center}
  \vspace{-0.5pc}
  \caption{RICH Counter structure (CAD view, left), matrix of detectors of the prototype in its 
testing environment (center), and detector cell used in the prototype including light guides, 
photomultiplier tube (PMT), read out electronics, and (half) plastic shell (right). 
\label{ARCHI}}
\end{figure}
The mechanical structure of the counter is shown on Fig.~\ref{ARCHI}. It consists of a radiator 
plane at the top, separated from the photodetector plane by a 45~cm drift space. 
The empty area in the detector plane corresponds to the location of the electromagnetic calorimeter.  
A conical mirror encloses the drift volume to increase the acceptance. The detector plane includes 
680~PMTs, corresponding to 10880 readout channels. The radiator plane will be equipped with NaF 
($n$=1.33) in the central region ($\sim$35$\times$35$\times$0.5~cm$^3$) and aerogel ($n$=1.03--1.05, 
3~cm thick) in the rest of the area.
\section{Prototype}
The prototype (Fig.~\ref{ARCHI}) consists of a fraction of the detector in a version close to the 
flight model design. This comprises 96 cells of photodetectors. Each cell includes a 16-anode PMT, 
16-element light guide matrix, HV dividers and front end readout electronics [1], mounted altogether 
and enclosed in a plastic shell (Fig.~\ref{ARCHI}), inserted in a magnetic shielding grid. The 
DAQ systems reads out 1536 channels in total. 
\section{The CERN SPS test beam}
The beam was obtained from the CERN SPS by fragmentation of primary Pb ions on a production
target located at the entrance of the beam line. The beam line was operated as a magnetic 
spectrometer and used to select samples of ions with a given rigidity. The fragmentation 
products fly to a good approximation at the beam velocity $\gamma_b$, and only those particles 
entering the beam line with a ratio $A/Z$ matching the rigidity settings for the beam line 
$B\rho=3.1 \gamma_b A/Z$, $\gamma_b$ beam Lorentz factor, were transported to the detector. 
The most useful ratio $A/Z$=2 provided a secondary beam including the whole range of nuclear 
masses: $^2$H,$^4$He,$^6$Li,$^{10}$B,$^{12}$C,$^{14}$N,$^{16}$O,..,$^{28}$Si,...,$^{40}$Ca,...,
$^{52}$Fe,..etc.., produced by projectile fragmentation. Other more selective field settings 
have been used to pick one $A/Z\neq$~2 isotope ratio, like 3/2 for $^3$He and 7/4 for $^7$Be.
Details on the beam design and performances are given in Ref.~[2].
\section{Results}
The prototype has been tested with both CRs (mainly muons) and beam ions. The instrumental environment 
was the same as used previously for the study prototype [3]. Only the in-beam test results are 
reported here (see [4] for the formers). The tests were performed over 4 days on october 2002, at 
CERN. Incident 20~GeV/$c$ per nucleon Pb ions were used to bombard a production target (10~cm Be 
or 40~cm Pb). The secondary beam intensity was set to about 1 to 3$\cdot$10$^3$ particles per spill. 
\begin{figure}[htb]
%\vspace{-0.5cm}
  \begin{center}
\hspace{-1cm}
    \includegraphics[height=4cm]{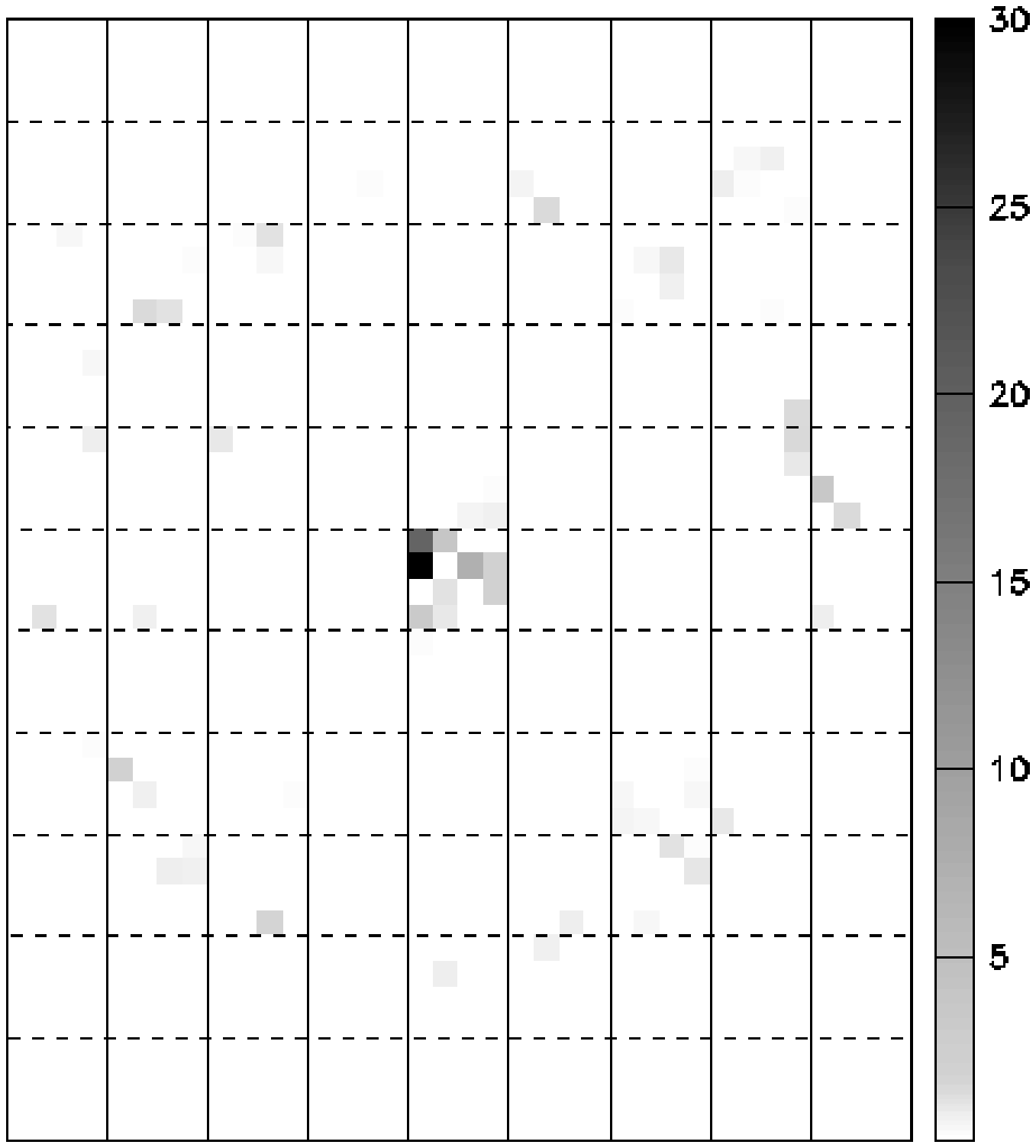}
\hspace{-0.5cm}    \includegraphics[height=4cm]{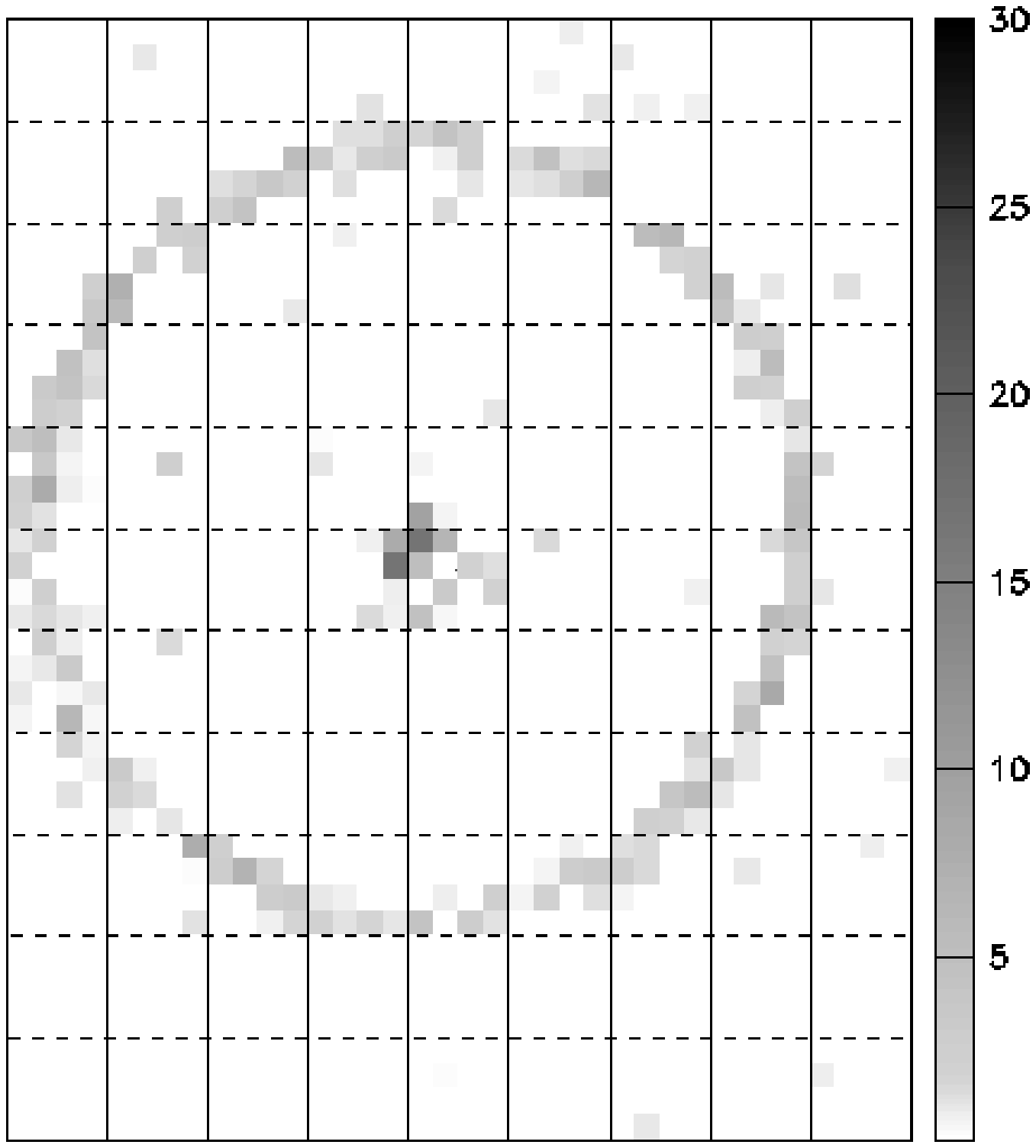}
\hspace{-0.5cm}    \includegraphics[height=4cm]{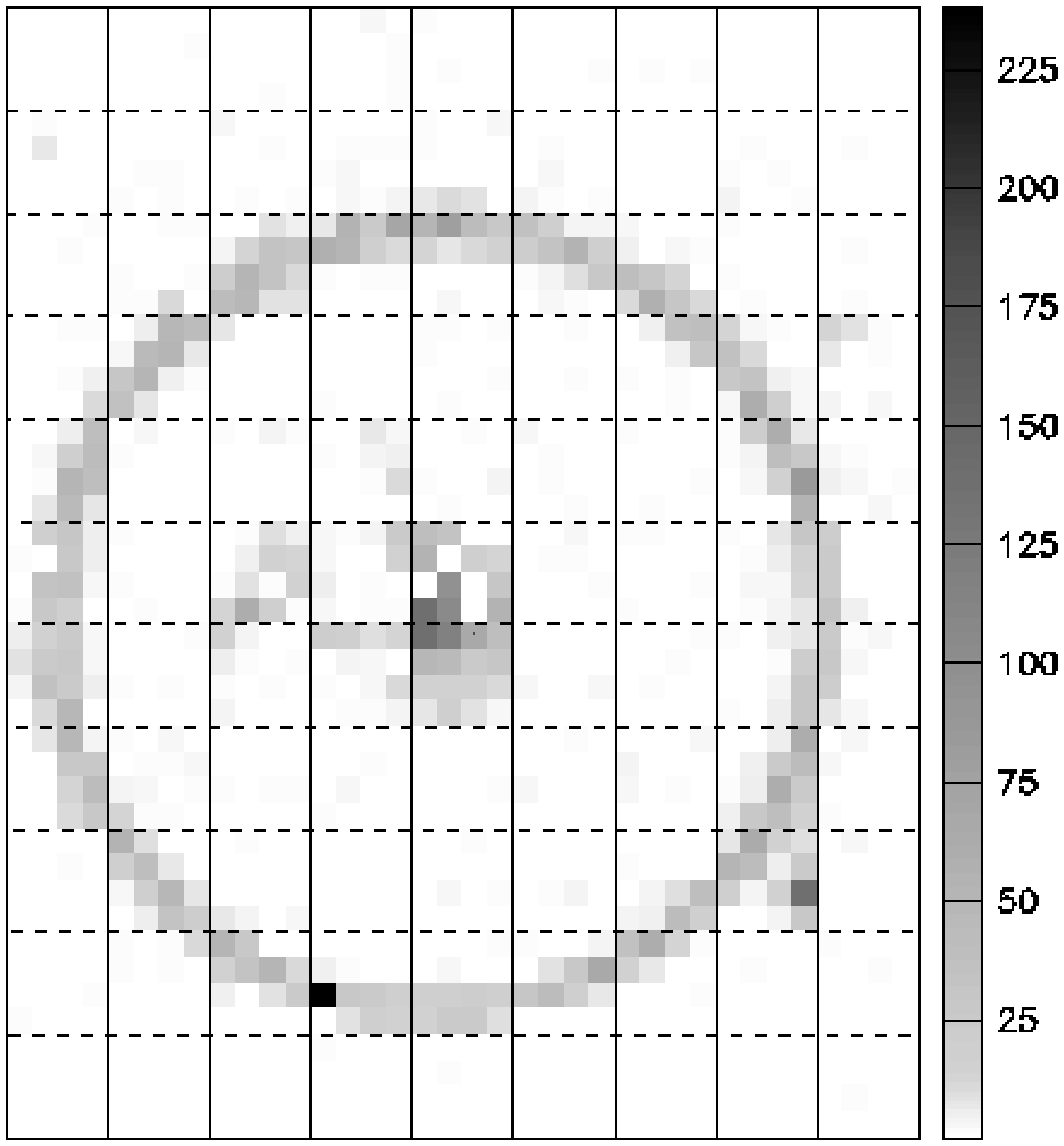}
\hspace{-0.5cm}    \includegraphics[height=4cm]{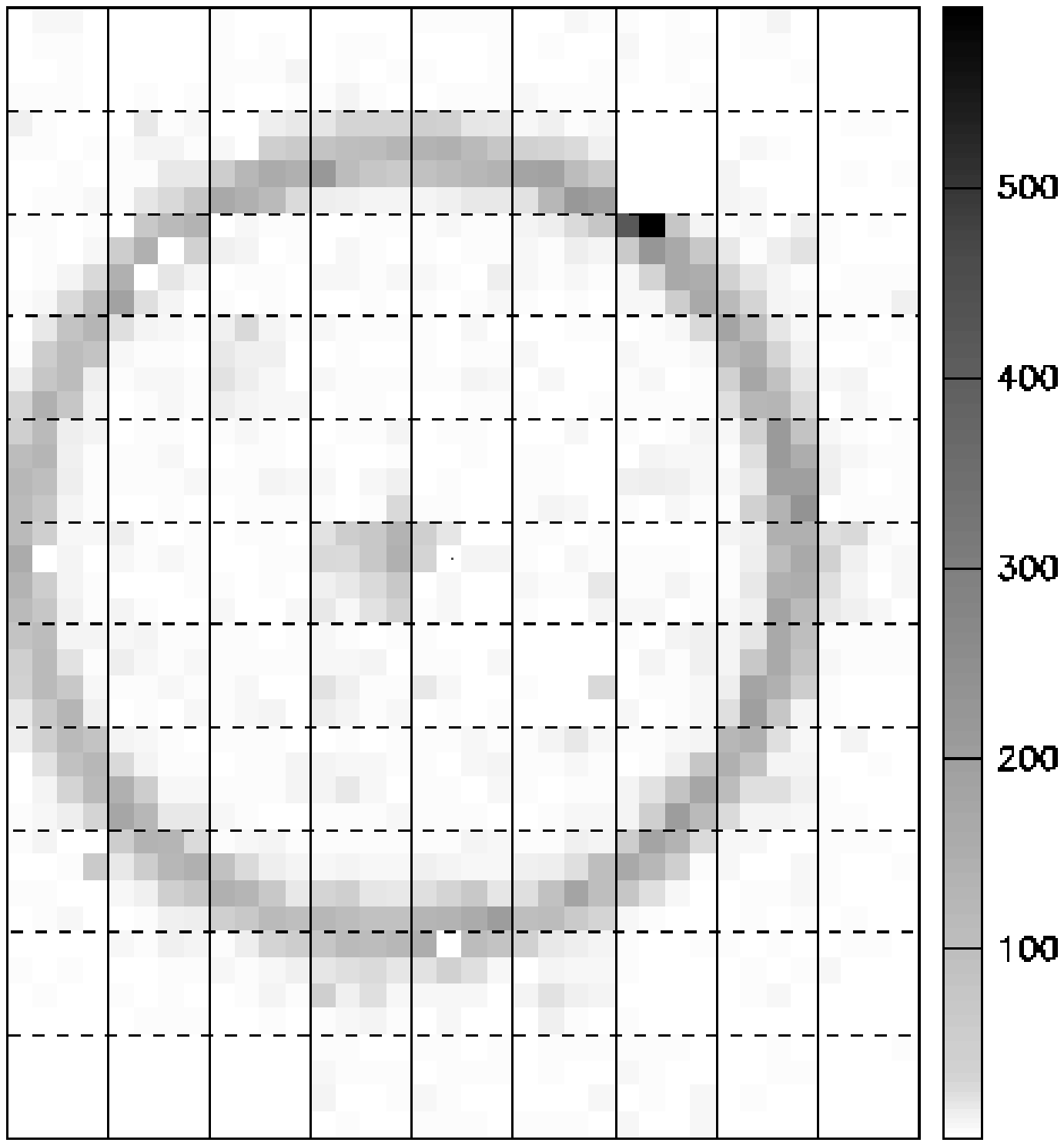} \\
    \includegraphics[width=6cm]{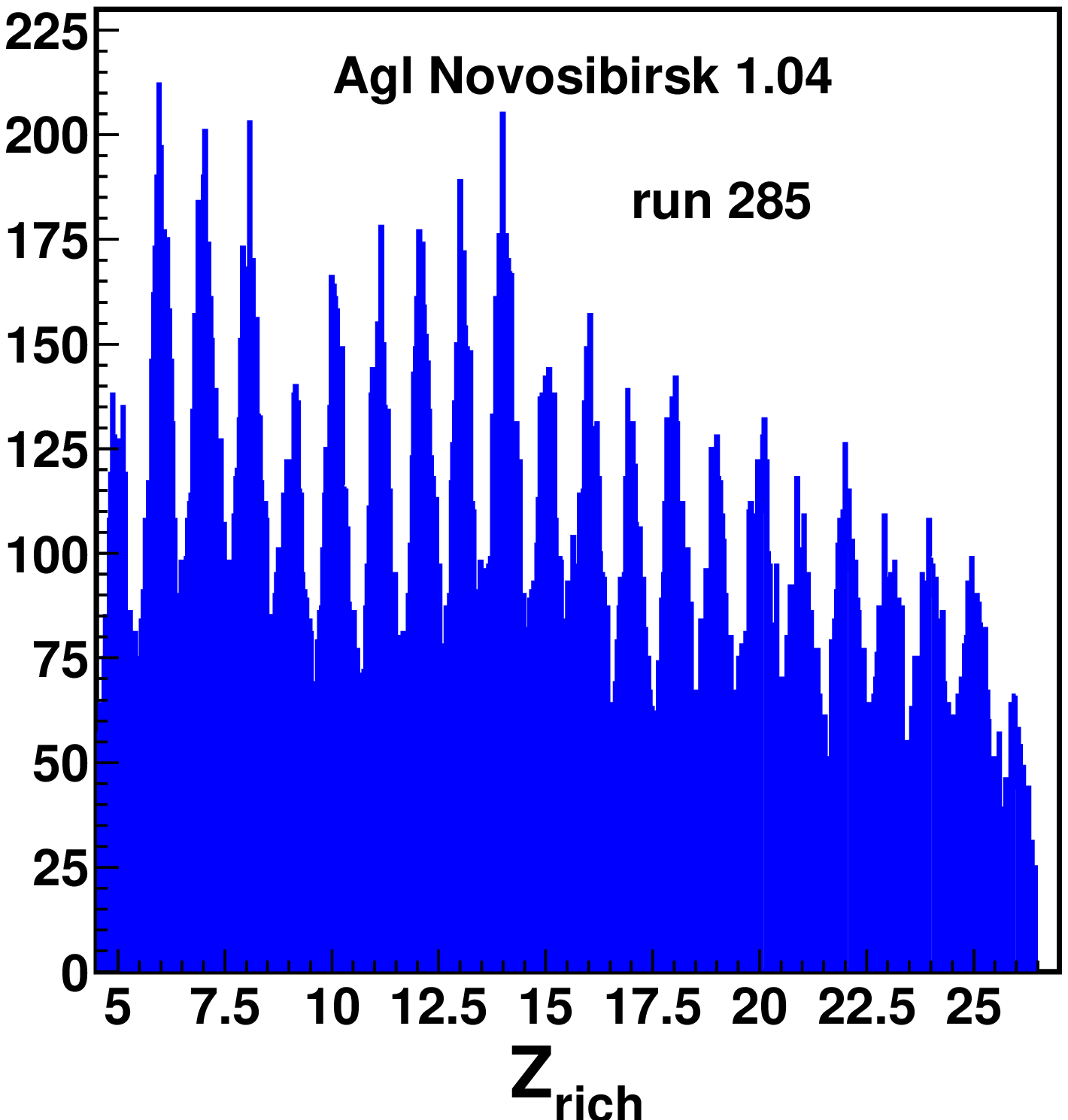}
    \includegraphics[width=6cm]{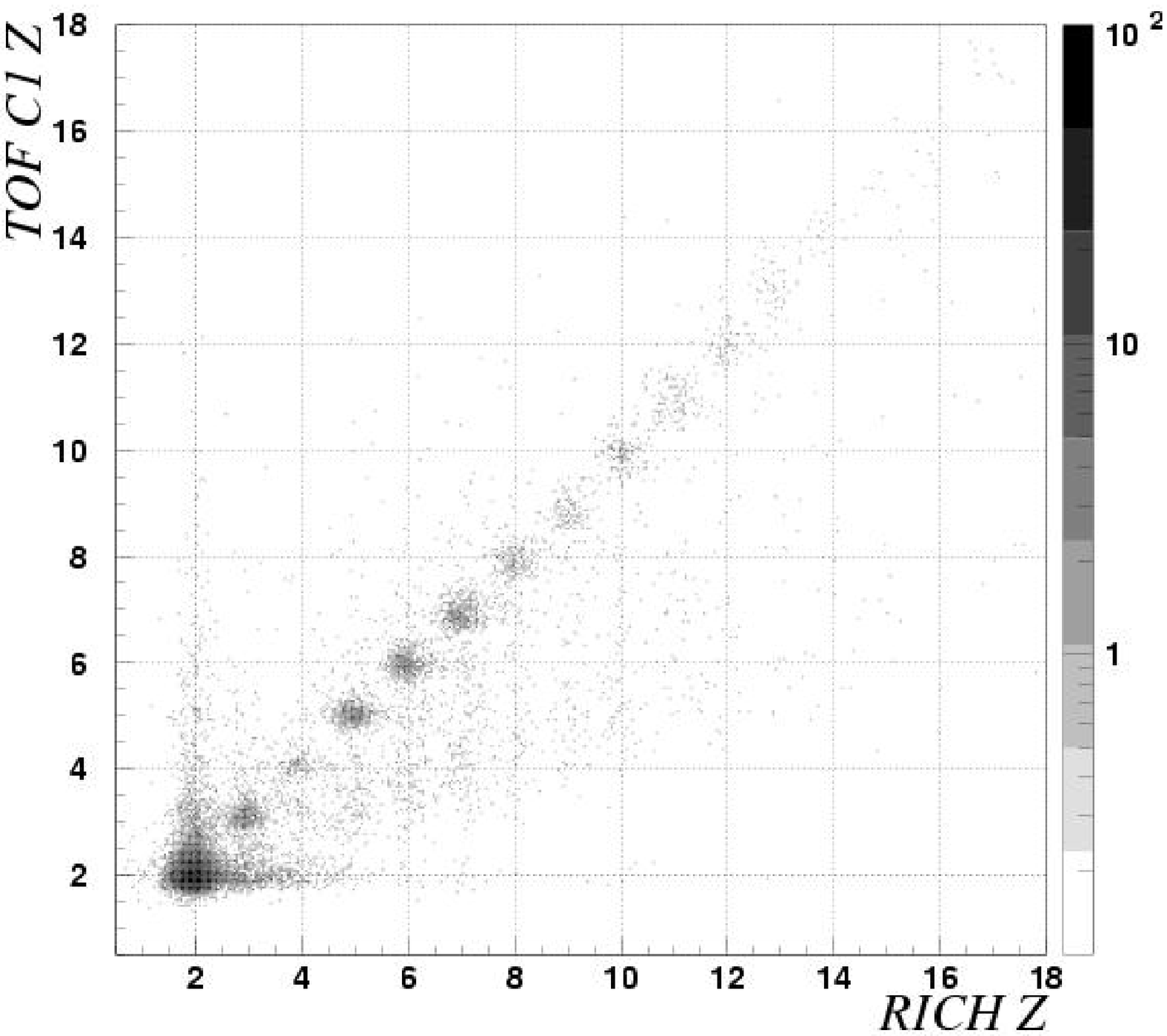}
  \end{center}
  \vspace{-0.5cm}
  \caption{Cherenkov patterns obtained with various ions with $Z$~=~2, 6, 16, and $>$25, 
from left to right respectively (top); $Z$ spectrum obtained from the RICH (bottom left), and 
correlation with the TOF measurement (bottom right) [5], measured with the ion beam at CERN. 
\label{RNGS}}
\end{figure}
The imaging performances of the prototype have been studied for various radiators: Silica aerogels 
with refraction index 1.03 to 1.05 from various manufacturers, and Sodium Fluoride (NaF). Cherenkov 
rings associated to ions over the covered range of charge have been observed, with the number of photons 
ranging from 4--7 (mean) for $Z$=1 elements, up to several thousands for high $Z$ values. The charge and 
velocity resolutions are being evaluated from the analysis of the 5 million events recorded. A realistic 
evaluation of the isotope separation capability of the final counter will be produced from this 
analysis.

Fig.~\ref{RNGS} shows a sample of results: Selection of single rings with different $Z$ values  
obtained with a 3~cm thick $n$=1.03 aerogel radiator (top); Preliminary results of analysis showing 
the prototype performance for $Z$ measurement with 3.1~cm thick $n=$1.04 aerogel (bottom left); And 
correlation with TOF hodoscopes d$E$/d$X$ measurements (bottom right, see [5]), illustrating the ID 
capability of the spectrometer for isotopes. The measured resolution in $Z$ is of the order of 
$\sigma(Z)$=~0.3 in the region of Fe ($Z$=26), in fair agreement with the estimates. %[*6].
Although the dynamic range goes beyond, and although higher $Z$ data have been recorded in the run, up 
to around $Z$=45, the study of this range suffers from the lack of good external $Z$ measurements. The 
work is in progress. 

The velocity ($\beta$) resolution was found in agreement with the CR estimates, i.e., around 
$\frac{\Delta\beta}{\beta}\sim 10^{-3}$ for $Z$=1 particles. For larger $Z$ values, the resolution 
first decreases (improves) as $Z^{-1}$ for $Z<$~7 as observed previously [3], then starts decreasing 
progressively slower than $Z^{-1}$ up to $Z\sim$~25, flattening above this value to a $Z$ independent, 
constant value, according to the relation $\sigma(\beta)=10^{-4}\sqrt{(8.2)^2/Z^2+ (1.2)^2}$. This 
latter limitation is set by the pixel size of the detector plane (evaluation on 1.03 aerogel data 
points). 

\vspace{\baselineskip}
\re
1.\ Eraud L., and Gallin-Martel L. \ 2002, Proc. Conf. on New Developments in Photodetection, 
June 17-21, 2002, Beaune, France, Nucl. Inst. and Meth. in Phys., to appear.
\re
2. Bu\'enerd M., and Efthymiopoulos I.\ 2003, Report CERN-AB-2003, May 2003, submitted. 
\re
3.\ Thuillier T. et al.\ 2002, Nucl. Inst. and Meth. in Phys. A491, 83.
\re
4.\ The AMS RICH collaboration (Casaus J. et al.) \ 2002, Nucl. Phys. B113, 147; \
 (Bu\'enerd M. et al.) \ 2003, Nucl. Inst. and Meth. in Phys. A502, 158.
\re
5.\ The AMS TOF collaboration (Casadei D. et al.), \ these proceedings.
%\re
%6.\ Bu\'enerd M., and Ren Z., \ 2000, Nucl. Inst. and Meth. in Phys. A454, 476

\endofpaper
\end{document}